\documentclass[aip,jcp,reprint]{revtex4-1}

\usepackage[utf8]{inputenc}
\usepackage{graphics,graphicx,epstopdf}
\usepackage{subfig}
\usepackage{dcolumn}
\usepackage{bm}
\usepackage{mathrsfs,amsmath}

\newcommand\p{\ensuremath{\partial}}

\newcommand\e{\epsilon}
\newcommand\F{\mathcal{F}}

\def\.{\cdot}

\begin{document}

\title{Control of chemical reactions using electric field gradients}

\author{Shivaraj D. Deshmukh}
\affiliation{Department of Chemical Engineering, Ben-Gurion University of the Negev, 
Beer-Sheva 84105, Israel}

\author{Yoav Tsori}
\email[]{tsori@bgu.ac.il}

\affiliation{Department of Chemical Engineering, Ben-Gurion University of the Negev, 
Beer-Sheva 84105, Israel}

\begin{abstract}

We examine theoretically a new idea for spatial and temporal control of chemical 
reactions. When chemical reactions take place in a mixture of solvents, an external 
electric field can alter the local mixture composition thereby accelerating or 
decelerating the rate of reaction. The spatial distribution of electric field strength 
can be non-trivial and depends on the arrangement of the electrodes producing it.
In the absence of electric field, the mixture is homogeneous and the reaction takes place 
uniformly in the reactor volume. When an electric field is applied the solvents separate 
and reactants are concentrated in the same phase or separate to different phases, 
depending on their relative miscibility in the solvents, and this can have a large effect 
on the kinetics of the reaction. This method could provide an alternative way to control 
runaway reactions and to increase the reaction rate without using catalysts.
\end{abstract}


\maketitle 

Chemical reactions are conventionally controlled by pressure, temperature, 
surface area, concentration, or by using a catalyst\cite{sanderson2011s,Atkins2006}. 
Attention has turned to active modification of molecular collision 
processes\cite{shapiro1992} and to the manipulation of activation energy 
barriers\cite{kohl2012,zocchi2010}.
Lasers were used for selectively making and breaking chemical 
bonds\cite{Assion1998} and current research looks at ways to exploit 
ultrafast lasers for mode-selective chemistry, stereodynamic, 
and quantum control of molecular processes\cite{shapiro1992,Zare1998,Sussman2006} in cost 
effective ways. Electric\cite{Tachiya1987,Tachiya2001,Kim2010} and magnetic fields and 
ultrasound are also used to control molecular collisions and thereby the 
chemistry\cite{Lemeshko2013,Brouard2014}. 
Electric and magnetic fields modify the orientation of 
molecules and may change the ion and molecular transport rates, they 
can modify the quantum states of molecules\cite{Lemeshko2013,Brouard2014} and they may 
lead to Stark and Zeeman shifts in the energy levels. 
In addition, large electric fields at charged metal surfaces are important in the 
chemisorption of atoms and molecules\cite{Kreuzer1991}.

Here we propose a new direction for spatial and temporal control of chemical reactions. 
Our idea relies on the use of a mixture of two or more solvents. Reactions taking place 
in such mixture are influenced by the composition of the solvents. An external field 
can lead to ``electro prewetting'' transition of the solvents even when their 
initial state is homogeneous \cite{Tsori2004,tsori_rmp2009}. In this transition, the 
solvents separate from each other and migrate to locations that minimize the total free 
energy of the mixture and depend on the electrode design. An interface thus appears 
separating the formed domains.

There are now two scenarios and for clarity we focus on a binary mixture of 
solvents and two reactants. In the first, due to their Gibbs transfer energy, the 
reactants are more miscible in the same solvent. In that case field-induced 
demixing will lead to concentration of the reactants in a small volume and to accelerated 
reaction kinetics compared to the no-field, homogeneous state. The product will be 
initially 
produced in the same small volume. In the second scenario, the two reactants are 
preferentially miscible in different solvents. In that case the reaction will take place 
at the interface separating liquid domains. This interfacial reaction is expected to be 
slowed down compared to the no-field (homogeneous) state. The product will initially be 
created at the interface between coexisting domains. In both cases, switching off the 
field will allow re-mixing of the solvents and return to the homogeneous state and to 
``normal'' kinetics.

Thermodynamics of liquid mixtures is underlying the reactions in electric fields. 
Consider for simplicity a mixture of two solvents. 
The total free energy of the mixture is given on the mean-field level as 
\begin{equation}
\F= \int_\Omega\left[ f_m(\phi,T) +f_e(\phi,{\bf E}) +f_i(\phi,T)\right] d\Omega~,  
\end{equation}
where $\Omega$ is the volume occupied by the mixture. The volume fraction of solvent $1$ 
is given by $\phi$ and the volume fraction of 
solvent $2$ is $1-\phi$, 
$T$ is temperature and $f_m$, $f_e$ and $f_i$ are mixing, electrostatic and 
interfacial energies. The mixing free energy per unit volume is given 
by\cite{Doi1996,safran1994} $f_m=(k_BT/v_0)\left[\phi \log \phi +(1-\phi) \log(1-\phi) 
+\chi \phi(1-\phi)\right]$,
where $k_B$ is the Boltzmann constant and $v_0$ is a molecular volume assumed here to be 
equal for  the two solvents.
$\chi$ is the Flory parameter varying inversely with $T$: $\chi\sim 1/T$. For positive 
values of $\chi$ the phase diagram has upper critical solution 
temperature\cite{Debye1965} and the critical point is given by 
$(\phi_c,\chi_c)=(1/2, 2)$. For such symmetric free energy expressions, the binodal curve 
$T_b(\phi)$ in the $\phi$-$T$ phase diagram is given by $\partial 
f(\phi,T_b)/\partial\phi=0$. Above this curve, the mixture is stable in the homogeneous 
phase. A quench to temperatures $T$ below $T_b$ leads to phase separation of two 
coexisting phases with compositions given by the values of the binodal at 
$T$. 

The interfacial free energy density is given by\cite{safran1994} 
$f_i=(k_BT/2v_0)\chi\lambda^2|\nabla\phi|^2$, 
where $\lambda$ is a constant related to interface width. 
The electrostatic free energy density is given by $f_e=-(1/2)\e_0 
\e(\phi)|\nabla\psi|^2$, where $\e_0$ is permittivity of free space, $\e$ is the relative 
permittivity of the mixture depending on its relative composition, and $\psi({\bf r})$ is 
the 
local electric potential.

In order for the electric field to be effective in separating the liquids, it must have 
large spatial gradients. We chose the simple ``wedge'' geometry to illustrate the
concept of chemical reactions. The wedge is comprised of two flat 
and conducting plates oriented with an opening angle $\beta$ between them 
($\beta=0^\circ$ corresponds to a parallel-plate capacitor). In this effectively 
two-dimensional system, the electric field depends only on the distance $r$ from the 
imaginary meeting point of the plates and is oriented in the azimuthal $\hat{\theta}$ 
direction\cite{jen2014}.

In the wedge geometry chosen, all quantities depend on $r$ alone and the hydrodynamic 
flow velocity ${\bf v}$ vanishes. The equations governing the demixing dynamics 
are then\cite{onuki2004,jen2014}
\begin{subequations} \label{eq:dynamics}
\begin{align} 
 \frac{\p\phi}{\p t} = D \nabla^2\frac{\delta f}{\delta 
\phi}~,~~~~ {\bf \nabla}\cdot[\e(\phi){\bf \nabla}\psi] =0~,
\end{align}
\end{subequations}
where $f=f_m+f_i+f_e$ is the total mixture free energy density and $D$ is an 
Onsager diffusivity constant taken here to be independent of $\phi$.

We consider a simple irreversible reaction $A+B\xrightarrow{k} 2C$, where 
molecules of compound $A$ and $B$ react to give product $C$ and $k$ is the rate constant. 
When the reaction takes place in a mixture of two solvents, each molecule $A$, $B$, and 
$C$ feels a spatially-dependent potential that depends on its relative solubility denoted 
by $u_a$, $u_b$, and $u_c$ respectively. For the $A$ molecule 
$u_a({\bf r})=u_1^a\phi+u_2^a(1-\phi)$ and similar expressions for 
$u_b({\bf r})$ and $u_c({\bf r})$. The parameters $u_1^a$ ad $u_2^a$ indicate
the solubilities of compound $A$ in liquids $1$ and $2$ respectively. The difference in 
the solubility parameters $\Delta u^a=u_2^a-u_1^a$ is related to the Gibbs transfer 
energy $\Delta 
G_t$ for transferring one $A$ molecule from a solvent with composition $\phi_1$ to 
a solvent with composition $\phi_2$ via $\Delta G_t=\Delta u^\pm(\phi_2-\phi_1)$. 
Experiments show that $G_t$ is on the order of $1-10k_BT$ in aqueous 
mixtures \cite{marcus_book1} and hence $\Delta u^\pm \sim 
1-10$.

The mass balance of compound $A$ gives the  modified reaction-diffusion equation: $\p C_a/\p t=D_a
\nabla^2 C_a + D_a\nabla \cdot \left[C_a \nabla( 
u_a/k_B T) \right]-k C_a C_b$. 
Here $D_a$ is the diffusion coefficient of compound $A$ in the mixture, assumed to be 
independent of $T$ and $\phi$. This equation can be recast in dimensionless form
\begin{equation}\label{MB_A}
\frac{\p \tilde{C}_a}{\p \tilde{t}} = \tilde{\nabla}^2 \tilde{C}_a + \tilde{\nabla} 
\cdot \left( \tilde{C}_a U'_a \tilde{\nabla} \phi \right) - \tilde{k}\tilde{C}_a 
\tilde{C}_b
\end{equation}
where $\tilde{C}_a=C_a/C_{a0}$ is the concentration scaled by $C_{a0}$, the initial 
(uniform) concentration of $A$ molecules, $\tilde{k}=k C_{a0}R_1^2/D_a$ is a scaled 
reaction rate, and $U'_a=(u_2^a-u_1^a)/k_B T$. The length is scaled using $R_1$ (the 
minimal distance $r$ from the imaginary meeting point of the plates): 
$\tilde{r}=r/R_1$, and time is scaled via $\tilde{t}=D_a/R_1^2t$. When a potential 
difference of magnitude $V$ is applied across wedge electrodes, the ratio of the
electrostatic energy stored in a molecular volume $v_0$ to the thermal energy is 
given by the dimensionless number\cite{Samin2009} $M_w\equiv 
V^2v_0\e_0/(4\beta^2k_BT_cR_1^2)$, where $T_c$ is critical temperature. Similar mass 
balance equation can be written for compounds $B$ and $C$:
\begin{eqnarray}\label{MB_B}
\frac{\p \tilde{C}_b}{\p \tilde{t}}&=&\frac{D_b}{D_a} \left[\tilde{\nabla}^2 \tilde{C}_b 
+ \tilde{\nabla} \cdot \left( \tilde{C}_b U'_b\tilde{\nabla} \phi \right)\right] - 
\tilde{k}
\tilde{C}_a \tilde{C}_b\\
\label{MB_C}
\frac{\p \tilde{C}_c}{\p \tilde{t}}&=&\frac{D_c}{D_a} \left[ \tilde{\nabla}^2 \tilde{C}_c 
+ \tilde{\nabla} \cdot \left( \tilde{C}_c U'_c\tilde{\nabla} \phi \right)\right] + 2 
\tilde{k} \tilde{C}_a \tilde{C}_b
\end{eqnarray}
with $\tilde{C}_b=C_b/C_{a0}$, $\tilde{C}_c=C_c/C_{a0}$, 
$U'_b=(u_2^b-u_1^b)/k_B T$ and $U'_c=(u_2^c-u_1^c)/k_B T$.

In this paper, we assume that the reaction is slow compared to the 
kinetics of phase separation. If this condition holds, the initially homogeneous mixture 
phase-separates on a fast time-scale to two coexisting domains with uniform densities of 
the reactants. The composition $\phi({\bf r})$ attains its equilibrium profile, 
minimizing the sum of mixture, interfacial, and electrostatic energies.
On a much longer scale, the chemical reaction then proceeds, with the 
compounds $A$, $B$, and $C$ experiencing a spatially-dependent force derivable from the 
$u$'s as is explained above. We further distinguish between the two cases whether the 
molecules $A$ and $B$ prefer the same phase or not.

\begin{figure}[!ht]
 \begin{center}
\includegraphics[width=3.5in]{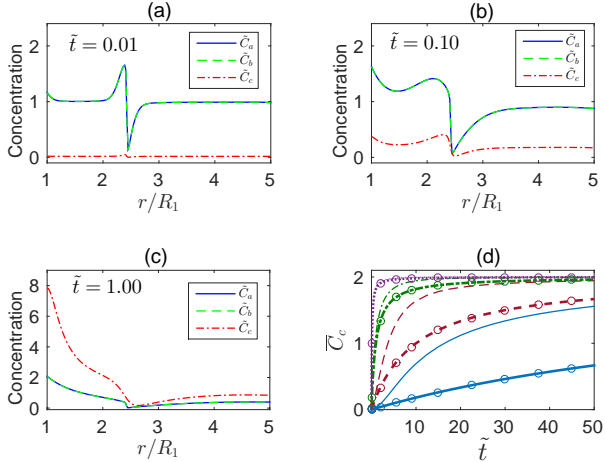} 
\end{center}
\caption{Concentration profiles $\tilde{C}_a$, $\tilde{C}_b$, and $\tilde{C}_c$ vs time 
when both $A$ and $B$ molecules are soluble in the polar phase. 
(a)-(c): temporal progression 
obtained from a numerical solution of Eqs. \ref{MB_A}, \ref{MB_B}, and \ref{MB_C}. 
(d): volume-average product $\bar{C}_c$ vs time for varying values of $\tilde{k}$. Solid, 
dashed, dash-dot and dotted lines correspond to $\tilde{k}=0.01$, $0.1$, $1$, $10$, 
respectively. Circles are the same but without field. 
The solubility potentials are $U'_a=U'_b=U'_c=10$, the diffusion constants are
$D_a=D_b=100\times D_c$, the average mixture composition is $\phi_0=0.33$, the 
dimensionless 
electric potential is $M_w=0.144$, and the 
dimensionless reaction rate is $\tilde{k}=1$.} 
\label{f1_WRM_CvsR_de5V20p0p33T0p962_dU10-dUc0_k1000Dc-11}
\end{figure}

\noindent {\bf Reactants $A$ and $B$ prefer the same phase.} 
Figure \ref{f1_WRM_CvsR_de5V20p0p33T0p962_dU10-dUc0_k1000Dc-11}(a)-(c) shows concentration 
profiles at various times when the reactants and product prefer the more polar phase (high 
value of $\phi$). At $\tilde{t}=0$ the mixture has two coexisting domains with an 
interface 
at $r/R_1\simeq 2.5$. As time progresses the $A$ and $B$ molecules diffuse towards the 
more polar region ($r/R_1<2.5$, large $\phi$) and sharp gradient occurs at the 
interface between the coexisting phases. The concentrations of $A$ and $B$ molecules is 
the 
same because we took $U'_a=U'_b$. As time increases, $A$ and $B$ continue to diffuse to 
the polar solvent but on the same time are consumed due to the reaction and the creation of 
$C$. At long times $\tilde{C}_a$ and $\tilde{C}_b$ become very small; the concentration of 
$C$ obeys the Boltzmann's distribution and is found preferentially in the polar phase 
where it is more soluble. 

In Figure \ref{f1_WRM_CvsR_de5V20p0p33T0p962_dU10-dUc0_k1000Dc-11}(d) we calculate the 
spatially-averaged product amount $\bar{C}_c\equiv \Omega^{-1}\int \tilde{C}_cd\Omega$ vs 
time. We compare the case with electric field 
and two coexisting domains (lines) with the zero-field case and a homogeneous mixture 
(lines with symbols), for constant electric field and varying reaction rates $\tilde{k}$. 
As 
can be seen, the electric field increases the effective rate of the reaction. 
It does so more effectively for the slow reactions (small values of $\tilde{k}$) and less 
effectively for the fast reactions (large values of $\tilde{k}$). Irrespective of 
$\tilde{k}$, material conservation dictates that $\bar{C}_c\to 2$ when 
$\tilde{t}\to\infty$. 

\begin{figure}[!ht]
\centering
\includegraphics[width=3in,viewport=0 -10 400 300]
{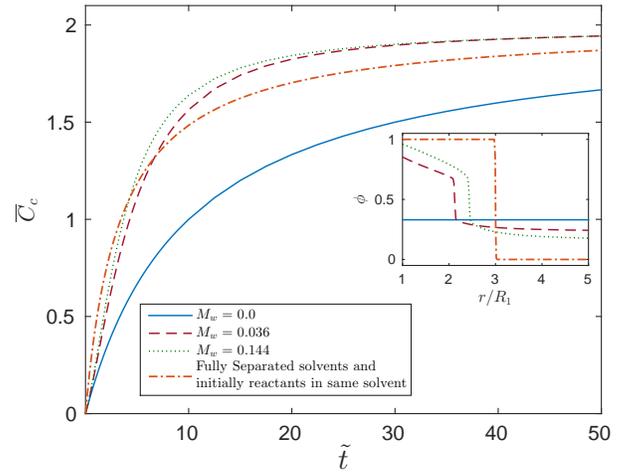} 
\caption{Dependence of the average product concentration $\bar{C}_c$ on the applied 
electric potential $M_w$. Inset shows the mixture profiles for the same $M_w$'s.  
$U'_a=U'_b=10$, $\phi_0=0.33$,  and $\tilde{k}=0.1$} 
\label{f2_WRM_AvgCcvst_de5VAllp0p33T0p962_dU10-dUc0_k100Dc-9}
\end{figure}

Fig. \ref{f2_WRM_AvgCcvst_de5VAllp0p33T0p962_dU10-dUc0_k100Dc-9} examines how the 
effective reaction rate depends on the magnitude of the external potential. 
A phase-separation transition occurs when the applied voltage in the wedge is larger than 
the critical value, which depends on the temperature and mixture composition. When the 
voltage increases past this threshold, the ``contrast'' between the phases increases 
($\phi$ increases at small values of $r$ and decreases for large values of $r$) and the 
interface displaces to larger values of $r$. Three profiles of $\phi$ for different 
values of the potential are shown in the inset. As can be seen, $\bar{C}_c$ is larger with 
increasing potential. However, an increase of $M_w$ from $0.036$ to $0.144$ has only a 
modest effect on $\bar{C}_c$. Indeed, even a hypothetical infinite potential would lead to 
a finite 
effective reaction rate.

\begin{figure}[!ht]
\centering
\includegraphics[width=3in]
{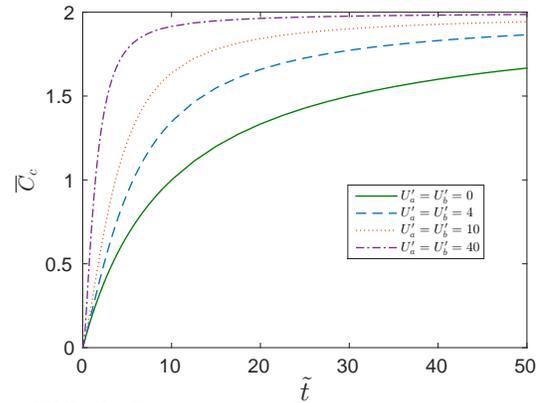} 
\caption{Dependence of the average product concentration $\bar{C}_c$ on the 
solubility parameters. 
$\phi_0=0.33$, $M_w=0.144$, and $\tilde{k}=0.1$}
\label{f3_WRM_AvgCcvst_de5V20p0p33T0p962_dUAll-dUc0_k100Dc-11}
\end{figure}

The preferential solubility of the reactants in the solvents is an important factor 
determining the rate of the reaction. In Fig. 
\ref{f3_WRM_AvgCcvst_de5V20p0p33T0p962_dUAll-dUc0_k100Dc-11} we plot $\bar{C}_c$ for 
different solubility values at a fixed value of $M_w$. While the asymptotic behavior at 
long times dictates $\bar{C}_c\to 2$ clearly, the dynamics are faster as the solubility 
difference increases.

\begin{figure}[!ht]
\begin{center}
\includegraphics[width=3.5in]{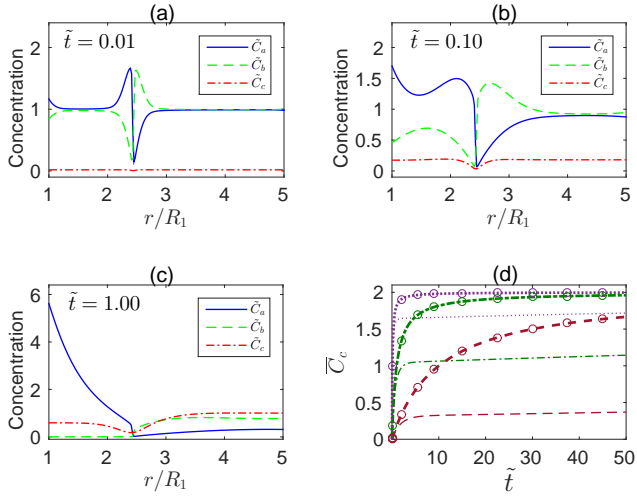} 
\end{center}
\caption{
Concentration profiles $\tilde{C}_a$, $\tilde{C}_b$, and $\tilde{C}_c$ and average 
product $\bar{C}_c$ for the same conditions as in Fig. 
\ref{f1_WRM_CvsR_de5V20p0p33T0p962_dU10-dUc0_k1000Dc-11} except that
molecule $A$ is soluble in the polar phase and $B$ in the less polar phase: 
$U'_a=10$ and $U'_b=-10$.} 
\label{f4_WRMd_CvsR_de5V20p0p33T0p962_dU10-dUc0_k1000Dc-11}
\end{figure}

\noindent
{\bf Reactants $A$ and $B$ prefer different phases.} 
We turn to the interesting case where the reactants are 
``antagonistic'' in the sense that they prefer different solvents.
Figure \ref{f4_WRMd_CvsR_de5V20p0p33T0p962_dU10-dUc0_k1000Dc-11} shows the concentration 
profiles vs time and should be compared to Fig. 
\ref{f1_WRM_CvsR_de5V20p0p33T0p962_dU10-dUc0_k1000Dc-11}. As before, $\tilde{C}_a$ 
and $\tilde{C}_b$ start from a uniform distribution at $\tilde{t}=0$. But here they 
diffuse in opposite directions -- $A$ to the polar solvent at small $r$'s and $B$ to the 
less polar solvent at large $r$'s. At early times the gradients in $\tilde{C}_a$ and 
$\tilde{C}_b$ occur at the interface between the solvents. The profiles evolve in time; 
as the reactants migrate according to their solubility they are consumed and $C$ is 
created. In this calculation we assumed the product is equally soluble in both solvents 
and thus at long times $\tilde{C}_c$ is uniform.
Far from the interface, in the bulk liquid, the transport of $A$ and $B$ 
is mainly due to diffusion (first term in the right-hand side of equations 
\ref{MB_A} and \ref{MB_B}). Near an interface the high gradient in $\phi$ leads to a 
large force from the solubility potential and 
transport is dominated by the solubility difference (second 
term in the right-hand side of equations \ref{MB_A} and \ref{MB_B}). 

\begin{figure}[!ht]
\centering
\includegraphics[width=3in]
{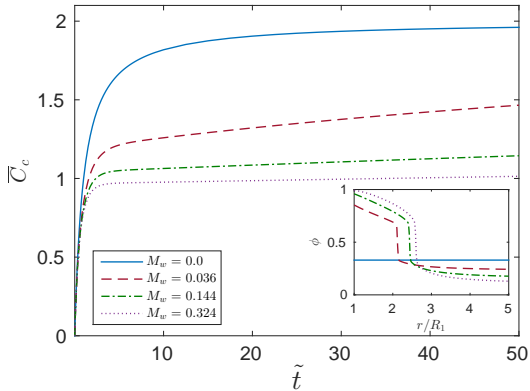} 
\caption{The influence of the electric potential $M_w$ on the average product 
concentration $\bar{C}_c$. Larger values of $M_w$ decrease the effective reaction rate. 
Inset shows the equilibrium profiles $\phi(r)$ for the same $M_w$'s. 
$U'_a=10$, $U'_b=-10$, $\phi_0=0.33$ and $\tilde{k}=1$.} 
\label{f5_WRMd_AvgCcvst_de5VAllp0p33T0p962_dU10-dUc0_k1000Dc-9}
\end{figure}

In Fig. \ref{f4_WRMd_CvsR_de5V20p0p33T0p962_dU10-dUc0_k1000Dc-11}(d) we plot the average 
product $\bar{C}_c$ vs time for different values of the potential. In contrast to Fig. 
\ref{f1_WRM_CvsR_de5V20p0p33T0p962_dU10-dUc0_k1000Dc-11}(d) here the effective reaction 
is {\it slower} with field (lines) as compared to the no-field case (lines with symbols). 
As in Fig. \ref{f1_WRM_CvsR_de5V20p0p33T0p962_dU10-dUc0_k1000Dc-11}(d), fast reactions 
are less affected by the field (large values of $\tilde{k}$). The total product $\tilde{C}_c$ 
tends to $2$ due to mass conservation.

The effective slowing-down of the reaction saturates with the magnitude of the applied 
potential, see Fig. \ref{f5_WRMd_AvgCcvst_de5VAllp0p33T0p962_dU10-dUc0_k1000Dc-9}. 
As $M_w$ increases ($M_w\propto V^2$) the more polar solvent is pulled to the region with 
higher electric 
field, thereby raising the value of $\phi$ at small $r$'s and reducing it at large $r$'s 
(low electric field region).
Curves show $\tilde{C}_c$ vs time for different values of $M_w$. All curves increase from 
$\tilde{C}_c=0$ at $\tilde{t}=0$ to $\tilde{C}_c=2$ at infinity. Curves exhibit an 
early increase on a small timescale dictated by diffusion over the mesoscopic width of 
the interface between coexisting phases. After the rapid increase, the curves increase 
slowly on a timescale dictated by diffusion over the macroscopic size of the wedge. 

\begin{figure}[!ht]
\centering
\includegraphics[width=3in]
{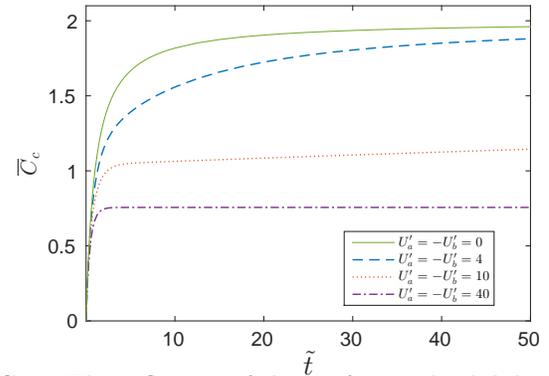} 
\caption{The influence of the preferential solubility of $A$ and $B$ molecules on the 
average total product $\bar{C}_c$. The reaction is considerably slowed down as $|U'_a|$ 
and $|U'_b|$ increase. $\phi_0=0.33$, $M_w=0.144$, and $\tilde{k}=1$.}
\label{f6_WRMd_AvgCcvst_de5V20p0p33T0p962_dUAll-dUc0_k1000Dc-9}
\end{figure}

The overall reaction rate is slowed down when the reactants are preferentially soluble in 
different solvents, as can be seen in Fig. 
\ref{f6_WRMd_AvgCcvst_de5V20p0p33T0p962_dUAll-dUc0_k1000Dc-9}. Clearly the electric 
potential is crucial here because in its absence the mixture is homogeneous and with it 
two domains exist. All curves in the Figure start from zero and increase to $2$ at long 
times. Here again there are two timescales -- a fast transient corresponding to diffusion 
over the interface, accompanied by a slow relaxation dictated by diffusion over 
macroscopic lengths. The stronger the incompatibility between the reactants the 
larger the difference in concentration of $A$ and $B$ molecules in the two domains 
and the slower is the overall relaxation.

\begin{figure}[!ht]
\centering
\includegraphics[width=3in]{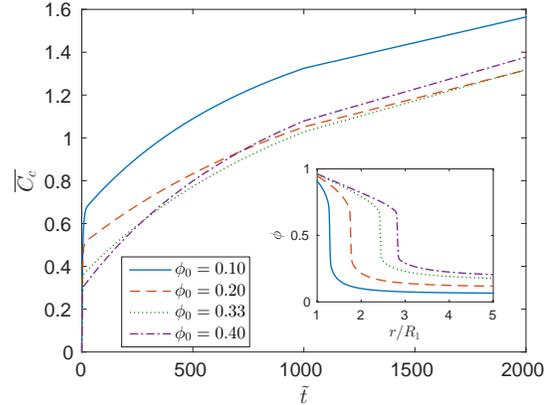}
\caption{Influence of the average mixture composition $\phi_0$ on the average total 
product $\bar{C}_c$ at a fixed electric potential given by $M_w=0.144$, solubilities 
$U'_a=-U'_b=10$ and $\tilde{k}=0.1$. Inset is the equilibrium profiles $\phi(r)$ for the 
same values of $\phi_0$.} 
\label{f7_WRMd_AvgCcvst_de5V20pAllT0p962_dU10-dUc0_k100}
\end{figure}

In Fig. \ref{f7_WRMd_AvgCcvst_de5V20pAllT0p962_dU10-dUc0_k100} we show 
the non-trivial dependence of the rate of reaction on the
average mixture composition $\phi_0$. At a fixed temperature and electric 
potential and for values of $\phi_0$ smaller than the critical value, an increase in 
$\phi_0$ to values closer to the binodal curve leads to an increase in the location of 
the interface between the polar and less polar phases (inset). However, at the same time 
the ``contrast'' between the phases diminishes and the thickness of the interface 
increases. As a result, for the values presented in this calculation, at long times the 
reaction which is slowest is one with intermediate value, $\phi_0=0.2$, and not one of 
the extreme values $\phi_0=0.1$ or $\phi_0=0.4$.

In summary, electric field gradients can control the effective reaction rate in 
reactions taking place in mixtures of solvents. Whether the reaction in the 'on' state is 
faster or slower than in the 'off' state depends on whether the reactants are miscible in 
the same solvent or in different ones.
The effective reaction rate depends sensitively on the electric 
potential, mixture composition, and distance from the binodal curve of the mixture, and 
is independent of the preferential solvation of the product $C$. 
We believe this method could provide an alternative way to control runaway reactions  
and to increase the reaction rate without using catalysts.

In the current study, we assumed that the liquid flow due 
to phase separation is very fast compared to the reaction kinetics but it may be 
interesting to relax this assumption. Electric fields are especially suited for use in 
microfluidic devices where they have been used to transport liquids in small channels. 
Such systems pose great promise for field-induced chemical reactions. Indeed, the challenge 
is to better understand how reactions are coupled with flow, either from pressure 
gradients in the channel or from the phase-separation itself. Future work will also 
consider ionic reagents, mixtures that have a Lower Critical Solution Temperature (e.g. 
Water and 2,6 Lutidine), \cite{tsori_pr_applied2014} and reversible reactions. In such 
reactions the preferential solubility of $C$ is important and the effective reaction rate 
can increase or decrease depending on whether $C$ is soluble in the same solvent as $A$ 
and $B$ or not.

This work was supported by the European Research Council ``Starting Grant'' No. 259205, 
COST Action MP1106, and Israel Science Foundation Grant No. 56/14.

\bibliography{draftWR}

\begin{thebibliography}{24}%
\makeatletter
\providecommand \@ifxundefined [1]{%
 \@ifx{#1\undefined}
}%
\providecommand \@ifnum [1]{%
 \ifnum #1\expandafter \@firstoftwo
 \else \expandafter \@secondoftwo
 \fi
}%
\providecommand \@ifx [1]{%
 \ifx #1\expandafter \@firstoftwo
 \else \expandafter \@secondoftwo
 \fi
}%
\providecommand \natexlab [1]{#1}%
\providecommand \enquote  [1]{``#1''}%
\providecommand \bibnamefont  [1]{#1}%
\providecommand \bibfnamefont [1]{#1}%
\providecommand \citenamefont [1]{#1}%
\providecommand \href@noop [0]{\@secondoftwo}%
\providecommand \href [0]{\begingroup \@sanitize@url \@href}%
\providecommand \@href[1]{\@@startlink{#1}\@@href}%
\providecommand \@@href[1]{\endgroup#1\@@endlink}%
\providecommand \@sanitize@url [0]{\catcode `\\12\catcode `\$12\catcode
  `\&12\catcode `\#12\catcode `\^12\catcode `\_12\catcode `\%12\relax}%
\providecommand \@@startlink[1]{}%
\providecommand \@@endlink[0]{}%
\providecommand \url  [0]{\begingroup\@sanitize@url \@url }%
\providecommand \@url [1]{\endgroup\@href {#1}{\urlprefix }}%
\providecommand \urlprefix  [0]{URL }%
\providecommand \Eprint [0]{\href }%
\providecommand \doibase [0]{http://dx.doi.org/}%
\providecommand \selectlanguage [0]{\@gobble}%
\providecommand \bibinfo  [0]{\@secondoftwo}%
\providecommand \bibfield  [0]{\@secondoftwo}%
\providecommand \translation [1]{[#1]}%
\providecommand \BibitemOpen [0]{}%
\providecommand \bibitemStop [0]{}%
\providecommand \bibitemNoStop [0]{.\EOS\space}%
\providecommand \EOS [0]{\spacefactor3000\relax}%
\providecommand \BibitemShut  [1]{\csname bibitem#1\endcsname}%
\let\auto@bib@innerbib\@empty
\bibitem [{\citenamefont {Sanderson}(2011)}]{sanderson2011s}%
  \BibitemOpen
  \bibfield  {author} {\bibinfo {author} {\bibfnamefont {K.}~\bibnamefont
  {Sanderson}},\ }\href@noop {} {\bibfield  {journal} {\bibinfo  {journal}
  {Nature}\ }\textbf {\bibinfo {volume} {469}},\ \bibinfo {pages} {18}
  (\bibinfo {year} {2011})}\BibitemShut {NoStop}%
\bibitem [{\citenamefont {Atkins}\ and\ \citenamefont
  {de~Paula}(2006)}]{Atkins2006}%
  \BibitemOpen
  \bibfield  {author} {\bibinfo {author} {\bibfnamefont {P.}~\bibnamefont
  {Atkins}}\ and\ \bibinfo {author} {\bibfnamefont {J.}~\bibnamefont
  {de~Paula}},\ }\href@noop {} {\emph {\bibinfo {title} {{Atkins' Physical
  Chemistry}}}}\ (\bibinfo  {publisher} {Oxford University Press},\ \bibinfo
  {address} {Oxford},\ \bibinfo {year} {2006})\BibitemShut {NoStop}%
\bibitem [{\citenamefont {Brumer}\ and\ \citenamefont
  {Shapiro}(1992)}]{shapiro1992}%
  \BibitemOpen
  \bibfield  {author} {\bibinfo {author} {\bibfnamefont {P.}~\bibnamefont
  {Brumer}}\ and\ \bibinfo {author} {\bibfnamefont {M.}~\bibnamefont
  {Shapiro}},\ }\href@noop {} {\bibfield  {journal} {\bibinfo  {journal} {Annu.
  Rev. Phys. Chem.}\ }\textbf {\bibinfo {volume} {43}},\ \bibinfo {pages} {257}
  (\bibinfo {year} {1992})}\BibitemShut {NoStop}%
\bibitem [{\citenamefont {Ratschbacher}\ \emph {et~al.}(2012)\citenamefont
  {Ratschbacher}, \citenamefont {Zipkes}, \citenamefont {Sias},\ and\
  \citenamefont {Kohl}}]{kohl2012}%
  \BibitemOpen
  \bibfield  {author} {\bibinfo {author} {\bibfnamefont {L.}~\bibnamefont
  {Ratschbacher}}, \bibinfo {author} {\bibfnamefont {C.}~\bibnamefont
  {Zipkes}}, \bibinfo {author} {\bibfnamefont {C.}~\bibnamefont {Sias}}, \ and\
  \bibinfo {author} {\bibfnamefont {M.}~\bibnamefont {Kohl}},\ }\href
  {http://dx.doi.org/10.1038/nphys2373} {\bibfield  {journal} {\bibinfo
  {journal} {Nat. Phys.}\ }\textbf {\bibinfo {volume} {8}},\ \bibinfo {pages}
  {649} (\bibinfo {year} {2012})}\BibitemShut {NoStop}%
\bibitem [{\citenamefont {Tseng}, \citenamefont {Wang},\ and\ \citenamefont
  {Zocchi}(2010)}]{zocchi2010}%
  \BibitemOpen
  \bibfield  {author} {\bibinfo {author} {\bibfnamefont {C.~Y.}\ \bibnamefont
  {Tseng}}, \bibinfo {author} {\bibfnamefont {A.}~\bibnamefont {Wang}}, \ and\
  \bibinfo {author} {\bibfnamefont {G.}~\bibnamefont {Zocchi}},\ }\href
  {http://stacks.iop.org/0295-5075/91/i=1/a=18005} {\bibfield  {journal}
  {\bibinfo  {journal} {Europhys. Lett.}\ }\textbf {\bibinfo {volume} {91}},\
  \bibinfo {pages} {18005} (\bibinfo {year} {2010})}\BibitemShut {NoStop}%
\bibitem [{\citenamefont {Assion}\ \emph {et~al.}(1998)\citenamefont {Assion},
  \citenamefont {Baumert}, \citenamefont {Bergt}, \citenamefont {Brixner},
  \citenamefont {Kiefer}, \citenamefont {Seyfried}, \citenamefont {Strehle},\
  and\ \citenamefont {Gerber}}]{Assion1998}%
  \BibitemOpen
  \bibfield  {author} {\bibinfo {author} {\bibfnamefont {A.}~\bibnamefont
  {Assion}}, \bibinfo {author} {\bibfnamefont {T.}~\bibnamefont {Baumert}},
  \bibinfo {author} {\bibfnamefont {M.}~\bibnamefont {Bergt}}, \bibinfo
  {author} {\bibfnamefont {T.}~\bibnamefont {Brixner}}, \bibinfo {author}
  {\bibfnamefont {B.}~\bibnamefont {Kiefer}}, \bibinfo {author} {\bibfnamefont
  {V.}~\bibnamefont {Seyfried}}, \bibinfo {author} {\bibfnamefont
  {M.}~\bibnamefont {Strehle}}, \ and\ \bibinfo {author} {\bibfnamefont
  {G.}~\bibnamefont {Gerber}},\ }\href
  {http://science.sciencemag.org/content/282/5390/919} {\bibfield  {journal}
  {\bibinfo  {journal} {Science}\ }\textbf {\bibinfo {volume} {282}},\ \bibinfo
  {pages} {919} (\bibinfo {year} {1998})}\BibitemShut {NoStop}%
\bibitem [{\citenamefont {Zare}(1998)}]{Zare1998}%
  \BibitemOpen
  \bibfield  {author} {\bibinfo {author} {\bibfnamefont {R.~N.}\ \bibnamefont
  {Zare}},\ }\href@noop {} {\bibfield  {journal} {\bibinfo  {journal}
  {Science}\ }\textbf {\bibinfo {volume} {279}},\ \bibinfo {pages} {1875}
  (\bibinfo {year} {1998})}\BibitemShut {NoStop}%
\bibitem [{\citenamefont {Sussman}\ \emph {et~al.}(2006)\citenamefont
  {Sussman}, \citenamefont {Townsend}, \citenamefont {Ivanov},\ and\
  \citenamefont {Stolow}}]{Sussman2006}%
  \BibitemOpen
  \bibfield  {author} {\bibinfo {author} {\bibfnamefont {B.~J.}\ \bibnamefont
  {Sussman}}, \bibinfo {author} {\bibfnamefont {D.}~\bibnamefont {Townsend}},
  \bibinfo {author} {\bibfnamefont {M.~Y.}\ \bibnamefont {Ivanov}}, \ and\
  \bibinfo {author} {\bibfnamefont {A.}~\bibnamefont {Stolow}},\ }\href
  {http://science.sciencemag.org/content/314/5797/278} {\bibfield  {journal}
  {\bibinfo  {journal} {Science}\ }\textbf {\bibinfo {volume} {314}},\ \bibinfo
  {pages} {278} (\bibinfo {year} {2006})}\BibitemShut {NoStop}%
\bibitem [{\citenamefont {Tachiya}(1987)}]{Tachiya1987}%
  \BibitemOpen
  \bibfield  {author} {\bibinfo {author} {\bibfnamefont {M.}~\bibnamefont
  {Tachiya}},\ }\href@noop {} {\bibfield  {journal} {\bibinfo  {journal} {J.
  Chem. Phys.}\ }\textbf {\bibinfo {volume} {87}},\ \bibinfo {pages} {4622}
  (\bibinfo {year} {1987})}\BibitemShut {NoStop}%
\bibitem [{\citenamefont {Hilczer}, \citenamefont {Traytak},\ and\
  \citenamefont {Tachiya}(2001)}]{Tachiya2001}%
  \BibitemOpen
  \bibfield  {author} {\bibinfo {author} {\bibfnamefont {M.}~\bibnamefont
  {Hilczer}}, \bibinfo {author} {\bibfnamefont {S.}~\bibnamefont {Traytak}}, \
  and\ \bibinfo {author} {\bibfnamefont {M.}~\bibnamefont {Tachiya}},\
  }\href@noop {} {\bibfield  {journal} {\bibinfo  {journal} {J. Chem. Phys.}\
  }\textbf {\bibinfo {volume} {115}},\ \bibinfo {pages} {11249} (\bibinfo
  {year} {2001})}\BibitemShut {NoStop}%
\bibitem [{\citenamefont {Reigh}, \citenamefont {Shin},\ and\ \citenamefont
  {Kim}(2010)}]{Kim2010}%
  \BibitemOpen
  \bibfield  {author} {\bibinfo {author} {\bibfnamefont {S.~Y.}\ \bibnamefont
  {Reigh}}, \bibinfo {author} {\bibfnamefont {K.~J.}\ \bibnamefont {Shin}}, \
  and\ \bibinfo {author} {\bibfnamefont {H.}~\bibnamefont {Kim}},\ }\href@noop
  {} {\bibfield  {journal} {\bibinfo  {journal} {J. Chem. Phys.}\ }\textbf
  {\bibinfo {volume} {132}},\ \bibinfo {pages} {164112} (\bibinfo {year}
  {2010})}\BibitemShut {NoStop}%
\bibitem [{\citenamefont {Lemeshko}\ \emph {et~al.}(2013)\citenamefont
  {Lemeshko}, \citenamefont {Krems}, \citenamefont {Doyle},\ and\ \citenamefont
  {Kais}}]{Lemeshko2013}%
  \BibitemOpen
  \bibfield  {author} {\bibinfo {author} {\bibfnamefont {M.}~\bibnamefont
  {Lemeshko}}, \bibinfo {author} {\bibfnamefont {R.~V.}\ \bibnamefont {Krems}},
  \bibinfo {author} {\bibfnamefont {J.~M.}\ \bibnamefont {Doyle}}, \ and\
  \bibinfo {author} {\bibfnamefont {S.}~\bibnamefont {Kais}},\ }\href
  {http://dx.doi.org/10.1080/00268976.2013.813595} {\bibfield  {journal}
  {\bibinfo  {journal} {Mol. Phys.}\ }\textbf {\bibinfo {volume} {111}},\
  \bibinfo {pages} {1648} (\bibinfo {year} {2013})}\BibitemShut {NoStop}%
\bibitem [{\citenamefont {Brouard}, \citenamefont {Parker},\ and\ \citenamefont
  {van~de Meerakker}(2014)}]{Brouard2014}%
  \BibitemOpen
  \bibfield  {author} {\bibinfo {author} {\bibfnamefont {M.}~\bibnamefont
  {Brouard}}, \bibinfo {author} {\bibfnamefont {D.~H.}\ \bibnamefont {Parker}},
  \ and\ \bibinfo {author} {\bibfnamefont {S.~Y.~T.}\ \bibnamefont {van~de
  Meerakker}},\ }\href {http://dx.doi.org/10.1039/C4CS00150H} {\bibfield
  {journal} {\bibinfo  {journal} {Chem. Soc. Rev.}\ }\textbf {\bibinfo {volume}
  {43}},\ \bibinfo {pages} {7279} (\bibinfo {year} {2014})}\BibitemShut
  {NoStop}%
\bibitem [{\citenamefont {Kreuzer}(1991)}]{Kreuzer1991}%
  \BibitemOpen
  \bibfield  {author} {\bibinfo {author} {\bibfnamefont {H.}~\bibnamefont
  {Kreuzer}},\ }\href
  {http://www.sciencedirect.com/science/article/pii/003960289190436V}
  {\bibfield  {journal} {\bibinfo  {journal} {Surf. Sci.}\ }\textbf {\bibinfo
  {volume} {246}},\ \bibinfo {pages} {336 } (\bibinfo {year}
  {1991})}\BibitemShut {NoStop}%
\bibitem [{\citenamefont {Tsori}, \citenamefont {Tournilhac},\ and\
  \citenamefont {Leibler}(2004)}]{Tsori2004}%
  \BibitemOpen
  \bibfield  {author} {\bibinfo {author} {\bibfnamefont {Y.}~\bibnamefont
  {Tsori}}, \bibinfo {author} {\bibfnamefont {F.}~\bibnamefont {Tournilhac}}, \
  and\ \bibinfo {author} {\bibfnamefont {L.}~\bibnamefont {Leibler}},\ }\href
  {http://www.nature.com/nature/journal/v430/n6999/abs/nature02758.html}
  {\bibfield  {journal} {\bibinfo  {journal} {Nature}\ }\textbf {\bibinfo
  {volume} {430}},\ \bibinfo {pages} {1} (\bibinfo {year} {2004})}\BibitemShut
  {NoStop}%
\bibitem [{\citenamefont {Tsori}(2009)}]{tsori_rmp2009}%
  \BibitemOpen
  \bibfield  {author} {\bibinfo {author} {\bibfnamefont {Y.}~\bibnamefont
  {Tsori}},\ }\href@noop {} {\bibfield  {journal} {\bibinfo  {journal} {Rev.
  Mod. Phys}\ }\textbf {\bibinfo {volume} {81}},\ \bibinfo {pages} {1471}
  (\bibinfo {year} {2009})}\BibitemShut {NoStop}%
\bibitem [{\citenamefont {Doi}(1996)}]{Doi1996}%
  \BibitemOpen
  \bibfield  {author} {\bibinfo {author} {\bibfnamefont {M.}~\bibnamefont
  {Doi}},\ }\href@noop {} {\emph {\bibinfo {title} {{Introduction to Polymer
  Physics}}}}\ (\bibinfo  {publisher} {Oxford University Press},\ \bibinfo
  {address} {Oxford},\ \bibinfo {year} {1996})\BibitemShut {NoStop}%
\bibitem [{\citenamefont {Safran}(1994)}]{safran1994}%
  \BibitemOpen
  \bibfield  {author} {\bibinfo {author} {\bibfnamefont {S.~A.}\ \bibnamefont
  {Safran}},\ }\href@noop {} {\emph {\bibinfo {title} {{Statistical
  Thermodynamics of Surfaces, Interfaces, And Membranes}}}}\ (\bibinfo
  {publisher} {Westview Press},\ \bibinfo {address} {New York},\ \bibinfo
  {year} {1994})\BibitemShut {NoStop}%
\bibitem [{\citenamefont {Debye}\ and\ \citenamefont
  {Kleboth}(1965)}]{Debye1965}%
  \BibitemOpen
  \bibfield  {author} {\bibinfo {author} {\bibfnamefont {P.}~\bibnamefont
  {Debye}}\ and\ \bibinfo {author} {\bibfnamefont {K.}~\bibnamefont
  {Kleboth}},\ }\href
  {http://link.aip.org/link/JCPSA6/v42/i9/p3155/s1{&}Agg=doi} {\bibfield
  {journal} {\bibinfo  {journal} {J. Chem. Phys.}\ }\textbf {\bibinfo {volume}
  {42}},\ \bibinfo {pages} {3155} (\bibinfo {year} {1965})}\BibitemShut
  {NoStop}%
\bibitem [{\citenamefont {Galanis}\ and\ \citenamefont
  {Tsori}(2014)}]{jen2014}%
  \BibitemOpen
  \bibfield  {author} {\bibinfo {author} {\bibfnamefont {J.}~\bibnamefont
  {Galanis}}\ and\ \bibinfo {author} {\bibfnamefont {Y.}~\bibnamefont
  {Tsori}},\ }\href@noop {} {\bibfield  {journal} {\bibinfo  {journal} {J.
  Chem. Phys.}\ }\textbf {\bibinfo {volume} {140}},\ \bibinfo {pages} {124505}
  (\bibinfo {year} {2014})}\BibitemShut {NoStop}%
\bibitem [{\citenamefont {Imaeda}, \citenamefont {Furukawa},\ and\
  \citenamefont {Onuki}(2004)}]{onuki2004}%
  \BibitemOpen
  \bibfield  {author} {\bibinfo {author} {\bibfnamefont {T.}~\bibnamefont
  {Imaeda}}, \bibinfo {author} {\bibfnamefont {A.}~\bibnamefont {Furukawa}}, \
  and\ \bibinfo {author} {\bibfnamefont {A.}~\bibnamefont {Onuki}},\
  }\href@noop {} {\bibfield  {journal} {\bibinfo  {journal} {Phys. Rev. E}\
  }\textbf {\bibinfo {volume} {70}},\ \bibinfo {pages} {051503} (\bibinfo
  {year} {2004})}\BibitemShut {NoStop}%
\bibitem [{\citenamefont {Marcus}(1985)}]{marcus_book1}%
  \BibitemOpen
  \bibfield  {author} {\bibinfo {author} {\bibfnamefont {Y.}~\bibnamefont
  {Marcus}},\ }\href@noop {} {\emph {\bibinfo {title} {{Ion Solvation}}}}\
  (\bibinfo  {publisher} {Wiley},\ \bibinfo {address} {New York},\ \bibinfo
  {year} {1985})\BibitemShut {NoStop}%
\bibitem [{\citenamefont {Samin}\ and\ \citenamefont
  {Tsori}(2009)}]{Samin2009}%
  \BibitemOpen
  \bibfield  {author} {\bibinfo {author} {\bibfnamefont {S.}~\bibnamefont
  {Samin}}\ and\ \bibinfo {author} {\bibfnamefont {Y.}~\bibnamefont {Tsori}},\
  }\href {http://www.ncbi.nlm.nih.gov/pubmed/19929041} {\bibfield  {journal}
  {\bibinfo  {journal} {J. Chem. Phys.}\ }\textbf {\bibinfo {volume} {131}},\
  \bibinfo {pages} {194102} (\bibinfo {year} {2009})}\BibitemShut {NoStop}%
\bibitem [{\citenamefont {Samin}\ \emph {et~al.}(2014)\citenamefont {Samin},
  \citenamefont {Hod}, \citenamefont {Melamed}, \citenamefont {Gottlieb},\ and\
  \citenamefont {Tsori}}]{tsori_pr_applied2014}%
  \BibitemOpen
  \bibfield  {author} {\bibinfo {author} {\bibfnamefont {S.}~\bibnamefont
  {Samin}}, \bibinfo {author} {\bibfnamefont {M.}~\bibnamefont {Hod}}, \bibinfo
  {author} {\bibfnamefont {E.}~\bibnamefont {Melamed}}, \bibinfo {author}
  {\bibfnamefont {M.}~\bibnamefont {Gottlieb}}, \ and\ \bibinfo {author}
  {\bibfnamefont {Y.}~\bibnamefont {Tsori}},\ }\href {<Go to
  ISI>://WOS:000344333400001} {\bibfield  {journal} {\bibinfo  {journal} {Phys.
  Rev. Appl.}\ }\textbf {\bibinfo {volume} {2}},\ \bibinfo {pages} {024008}
  (\bibinfo {year} {2014})}\BibitemShut {NoStop}%
\end{thebibliography}%
\end{document}